\documentclass{PoS}
\DeclareRobustCommand{\ion}[2]{%
\relax\ifmmode
\ifx\testbx\f@series
{\mathbf{#1\,\mathsc{#2}}}\else
{\mathrm{#1\,\mathsc{#2}}}\fi
\else\textup{#1\,{\mdseries\textsc{#2}}}%
\fi}
\title{Data reduction strategy of the Effelsberg-Bonn HI Survey (EBHIS)}

\ShortTitle{Data reduction strategy of the EBHIS}

\author{\speaker{B.~Winkel}, J.~Kerp and P.M.W.~Kalberla\\
        Argelander-Institut f\"{u}r Astronomie, University of Bonn\\
        E-mail: \email{bwinkel@astro.uni-bonn.de}}

\abstract{Since autumn 2008 a new L-band 7-Feed-Array receiver is used for an \ion{H}{i} 21-cm line survey performed with the 100-m Effelsberg telescope. The survey will cover the whole northern hemisphere comprising both, the galactic and extragalactic sky in parallel. Using state-of-the-art FPGA based digital Fast Fourier Transform spectrometers, superior in dynamic range and temporal resolution, allows to apply sophisticated radio frequency interferences (RFI) mitigation schemes to the survey data.

The EBHIS data reduction software includes the RFI mitigation, gain-curve correction, intensity calibration, stray-radiation correction, gridding, and  source detection. We discuss the severe degradation of radio astronomical \ion{H}{i} data by RFI signals and the gain in scientific yield when applying modern RFI mitigation schemes. For this aim simulations of the galaxy distribution within the local volume (z<0.07) with and without RFI degradation were performed. These simulations, allow us to investigate potential biases and selection effects introduced by the data reduction software and the applied source parametrization methods.   }

\FullConference{Panoramic Radio Astronomy: Wide-field 1-2 GHz research on galaxy evolution - PRA2009\\
		 June 02 - 05 2009\\
		 Groningen, the Netherlands}

\begin{document}

\section{Introduction}
Performing blind \ion{H}{i} surveys is a major challenge with respect to a variety of aspects. Modern receiving systems offer  total bandwidths of several hundreds MHz or even GHz, while digital high-dynamic-range  backends on the basis of  field programmable gate arrays (FPGA) allow to dump spectra on time-scales of seconds. The combination of both properties provides substantial improved data quality but leads to a much larger data volume. Reducing and analyzing the data is impossible to accomplish manually. 

Here, we present the data reduction strategy which is applied for the Effelsberg-Bonn \ion{H}{i} Survey (Section\,\ref{secreduction}). We performed simulations to test the software and investigate the influence of radio frequency interference (RFI) on the data (Section\,\ref{secsimulations}). The results give us the opportunity to identify systematic effects produced by the data reduction procedure itself (Section\,\ref{secsystematiceffects}). 

\section{The data reduction scheme}\label{secreduction}

With respect to \textit{RFI} the situation at telescope sites is getting worse, due to the increasing number of terrestrial radio wavelength applications across the Earth. RFI mitigation has to be treated right at the beginning of any data reduction chain, because interferences affect each reduction step significantly. There are different approaches, from real-time applications, e.g.,  adaptive filters 
or higher-order statistical analysis, 
 to various software-based solutions, e.g., simple (manual) flagging of polluted spectral channels, or more sophisticated methods as described in Winkel et~al. (2007 \cite{winkel07a}), the latter will be used for the EBHIS.

Another major issue is the \textit{gain curve calibration}. Using large bandwidths one often experiences strong ripples in the baseline produced by standing waves between the dish and the receiver. In combination with rather complex IF gain characteristics the bandpass calibration can be complicated. One solution is the implementation of the least-squares frequency switching (LSFS) method recently proposed by Heiles (2007 \cite{heiles07}), which needs only minors changes of the hardware in the IF processing chain (not yet implemented at the 100-m telescope). Unfortunately, LSFS fails in the presence of RFI signals, but Winkel \& Kerp (2007 \cite{winkel07b}) identified a workaround using flagging of polluted data points.

Very important for galactic astronomy is the correction for \textit{stray-radiation} (SR) --- emission entering via the side lobes of the antenna, which can seriously impact the measured fluxes. For EBHIS the method of Kalberla et~al. (2005 \cite{kalberla05}) is used. If the antenna diagram is well-known, the measured data can be corrected for the influence of the side lobe pattern to a good precision.

After RFI detection the data have to be \textit{calibrated} in terms of brightness temperature. A two-step method is used, first applying an absolute calibration via measuring a calibration source of well-defined flux density (S7), and then relative changes are determined using a noise diode the output of which is fed into the receiver during certain intervals. 

Finally, a graphical user interface (GUI) was developed for EBHIS  especially suited for the \textit{search and parametrization of sources} in the data cubes of the extragalactic part of the \ion{H}{i} survey. A finder algorithm based on the Gamma test was implemented (Boyce 2003 \cite{boyce03}). The GUI is designed to allow a fast working flow and is able to compute statistical errors of the fitted parameters using Markov-Chain Monte-Carlo methods.
\begin{figure}[!t]
\centering
\includegraphics[width=0.65\textwidth]{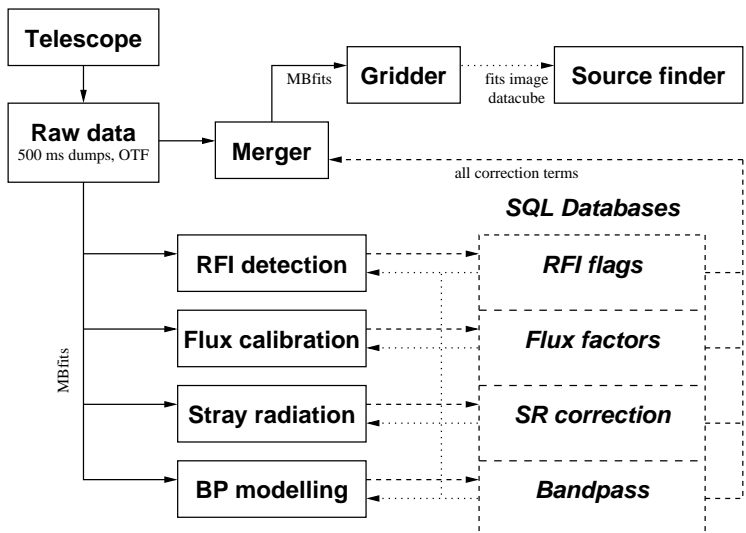}\\[-1ex]
\caption{Data reduction scheme used for EBHIS.}\label{figdatareductionscheme}
\end{figure}

The spectroscopic data are stored in the \texttt{MBfits} file format. The integration time per dump is $500\,\mathrm{ms}$, hence, a huge amount of data has to be handled for the full survey. The overall data processing is organized in a highly flexible way. To avoid  redundancy,  each module (RFI detection, SR correction, flux and bandpass calibration) works independently on the data and all correction terms are stored within a database, instead of producing complete sets of processed spectra after each intermediate step. For example, the RFI detection algorithm returns only the list of spectral channels per dump containing an RFI event. Therefore, only this list needs to be stored and can be directly used by the other tasks to flag bad data. The same principle holds for the flux calibration, where only one correction factor per spectrum must be present. It is possible to improve one of the modules without the  necessity to re-execute the other reduction tasks. A schematic view of the EBHIS data processing is shown in Fig.\,\ref{figdatareductionscheme}. In contrast to pipeline approaches, a new task is necessary, which is denoted as ``merger'', to apply all the correction terms to generate the final dataset.   Then the spectra are gridded to a data cube. 
To speed up computation, many tasks make use of multi-processor or -core platforms. Further improvement on processing speed is gained, because the database approach allows simultaneous workstation access. It also provides the opportunity to easily query useful informations like sky areas already observed (which can easily be visualized) and makes backups easier and smaller.

\section{Simulations}\label{secsimulations}
We carried out simulations to analyze the performance of the data reduction software in terms of source detection probability and data quality. This empirical approach  was chosen, because it provides the possibility of a quantitative analysis of source parameters and typical processing times. Also, the free parameters of the source detection algorithm could be tested. Another aim of the simulations was to study the influence of RFI signals on the data reduction quality and derived physical parameters. The synthetic spectra, therefore, have to contain simulated RFI events.

The simulations comprise the generation of artificial spectra, containing emission lines, noise, and eventually narrow-band RFI signals with time-variable amplitudes (randomly drawn from a power law distribution with exponent~$-1.5$). Our RFI detection algorithm was used to build up a \textit{flag database}. We computed three different sets of spectra, one set without interferences added (\textit{run 1}), the remaining two including RFI, but one without (\textit{run 2}) and one with RFI mitigation applied (\textit{run 3}). The data were gridded and the source detection and parametrization was completely automated (using our \textit{Galaxy Parametrizer} software). The obtained results can be directly compared to the simulated galaxy properties, leading to empirical measures of the data reduction quality.



One of the key properties of the local universe is the \ion{H}{i} mass function (HIMF), the number density of galaxies per mass interval and unit volume. The synthetic source sample was generated, such that it follows the HIMF as determined by HIPASS (Zwaan et~al. 2005 \cite{zwaan05}) .



\begin{figure}[!t]
\centering
\includegraphics[scale=0.48,bb=77 47 400 274,clip=]{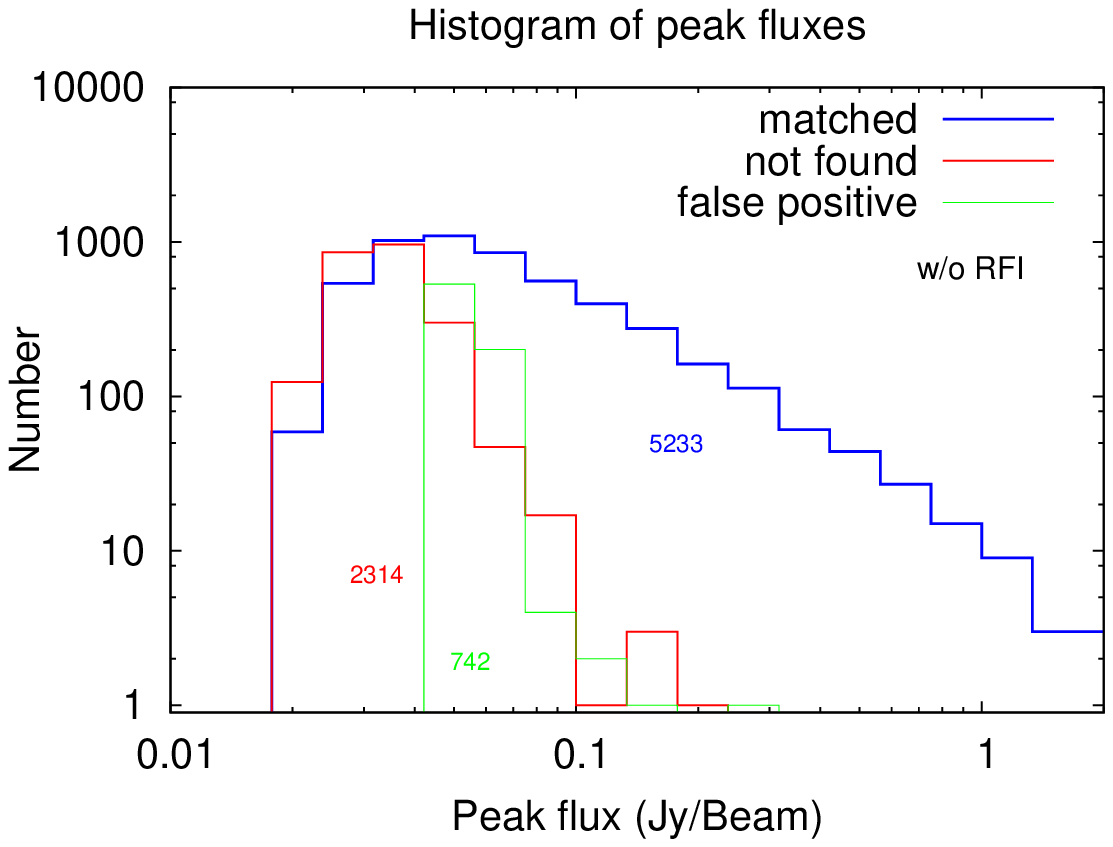}\raise8.5em\vbox{\moveleft11em\hbox{\small \it run (1)}}%
\includegraphics[scale=0.48,bb=126 47 400 274,clip=]{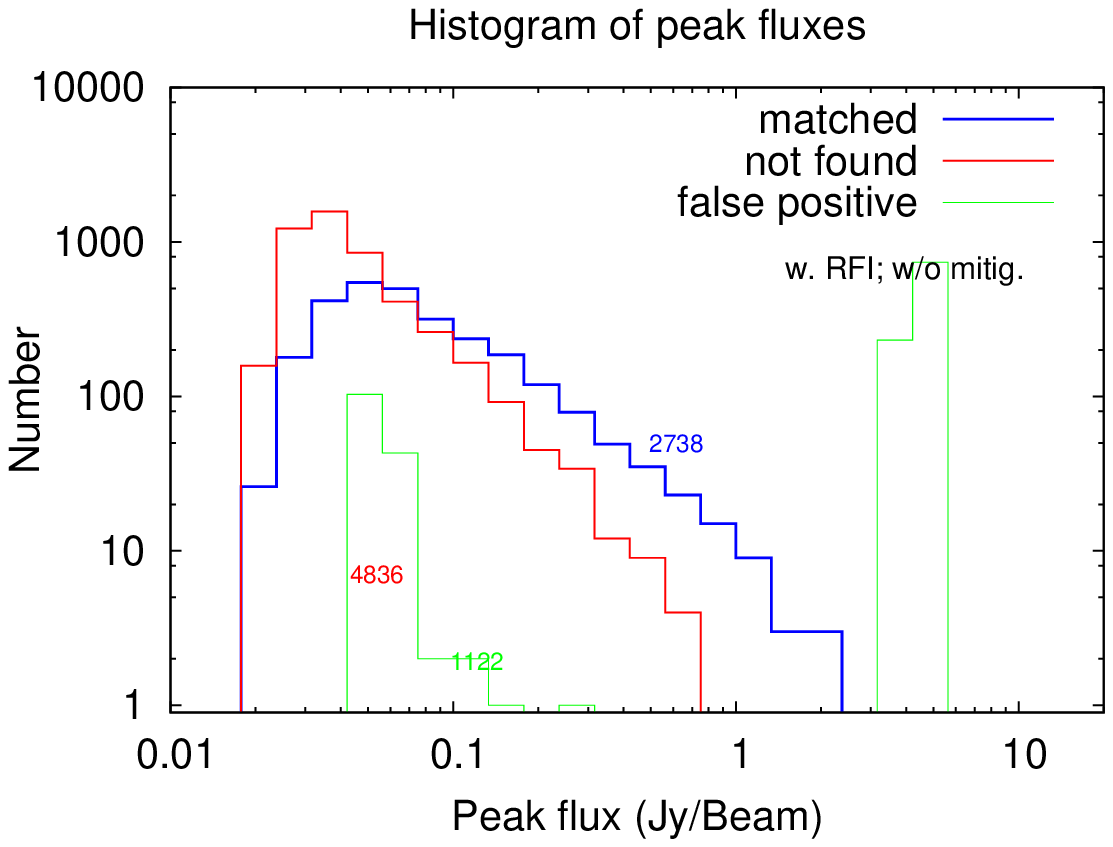}\raise8.5em\vbox{\moveleft11em\hbox{\small \it run (2)}}%
\includegraphics[scale=0.48,bb=126 47 400 274,clip=]{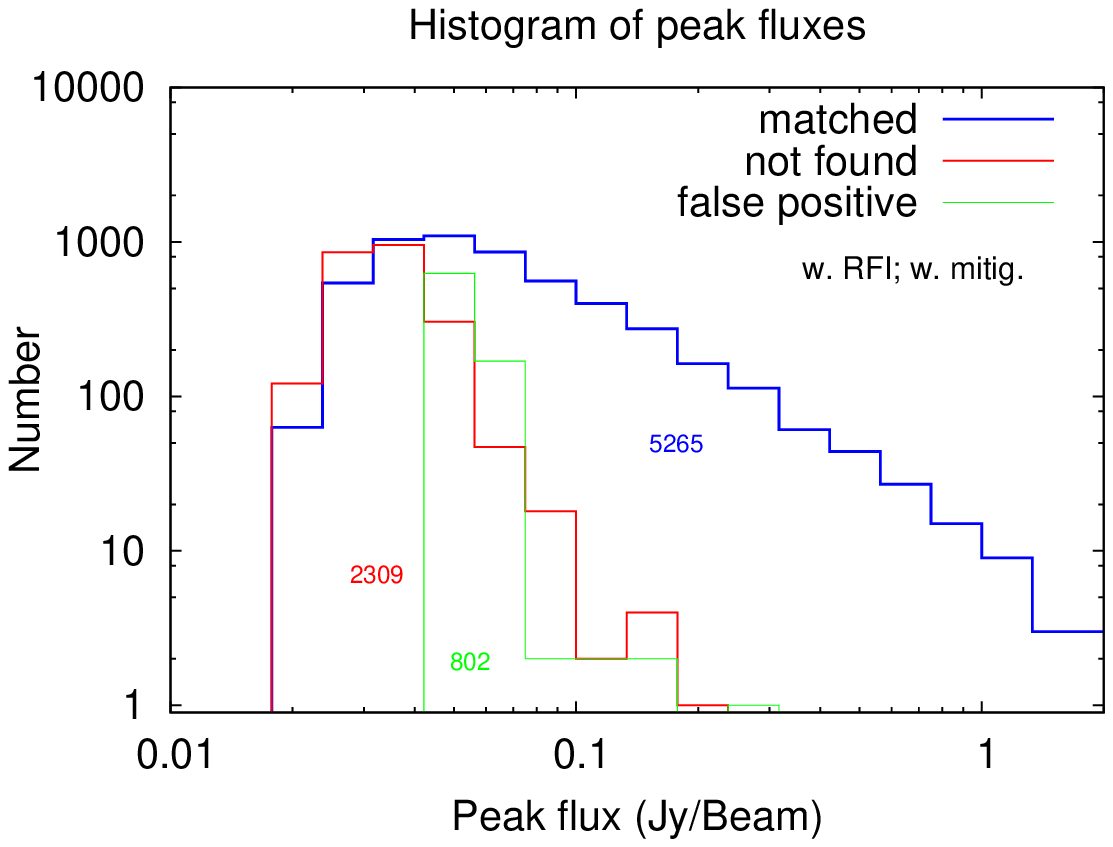}\raise8.5em\vbox{\moveleft11em\hbox{\small \it run (3)}}\\[0ex]
\includegraphics[scale=0.48,bb=77 47 400 274,clip=]{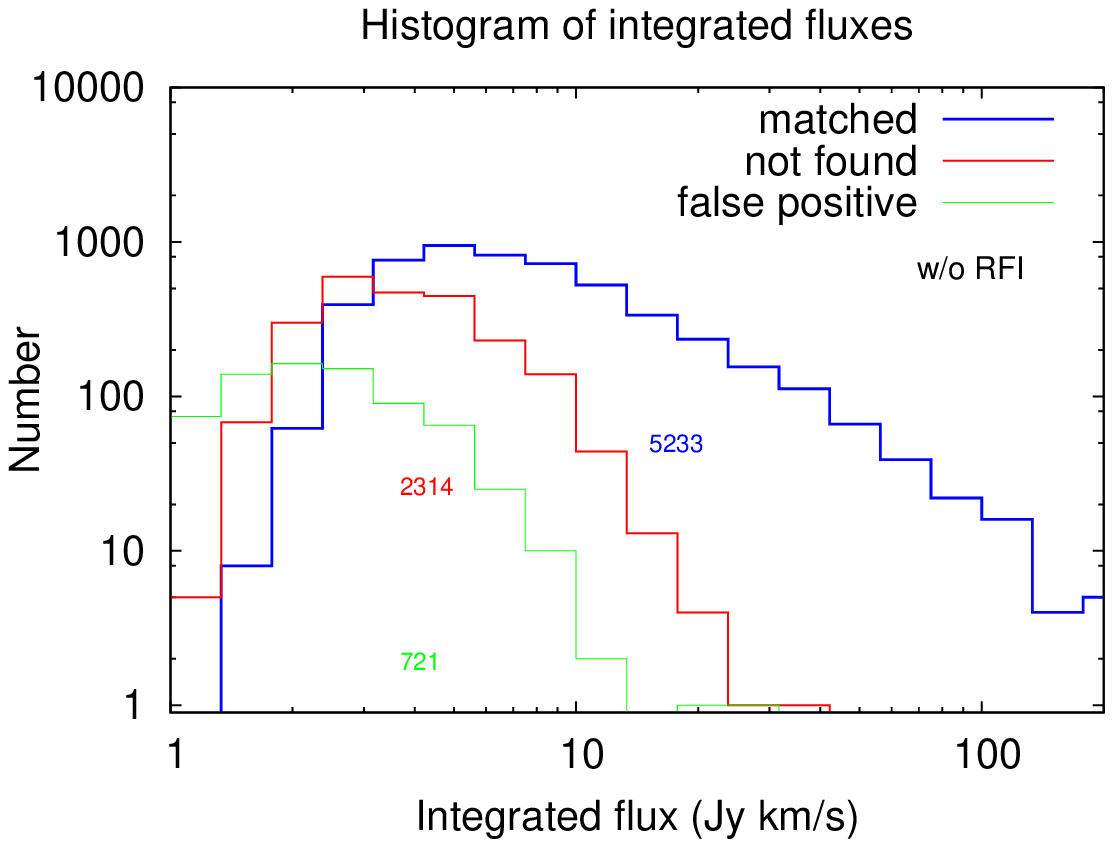}\raise8.5em\vbox{\moveleft11em\hbox{\small \it run (1)}}%
\includegraphics[scale=0.48,bb=126 47 400 274,clip=]{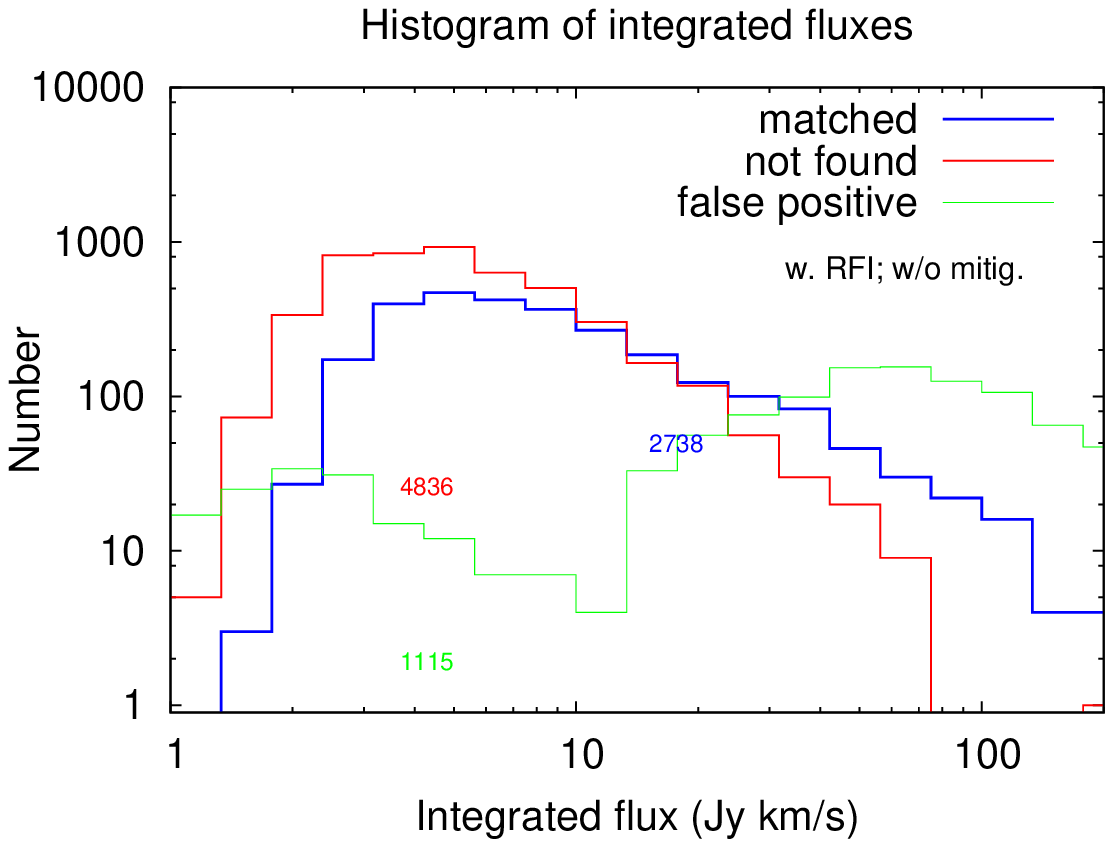}\raise8.5em\vbox{\moveleft11em\hbox{\small \it run (2)}}%
\includegraphics[scale=0.48,bb=126 47 400 274,clip=]{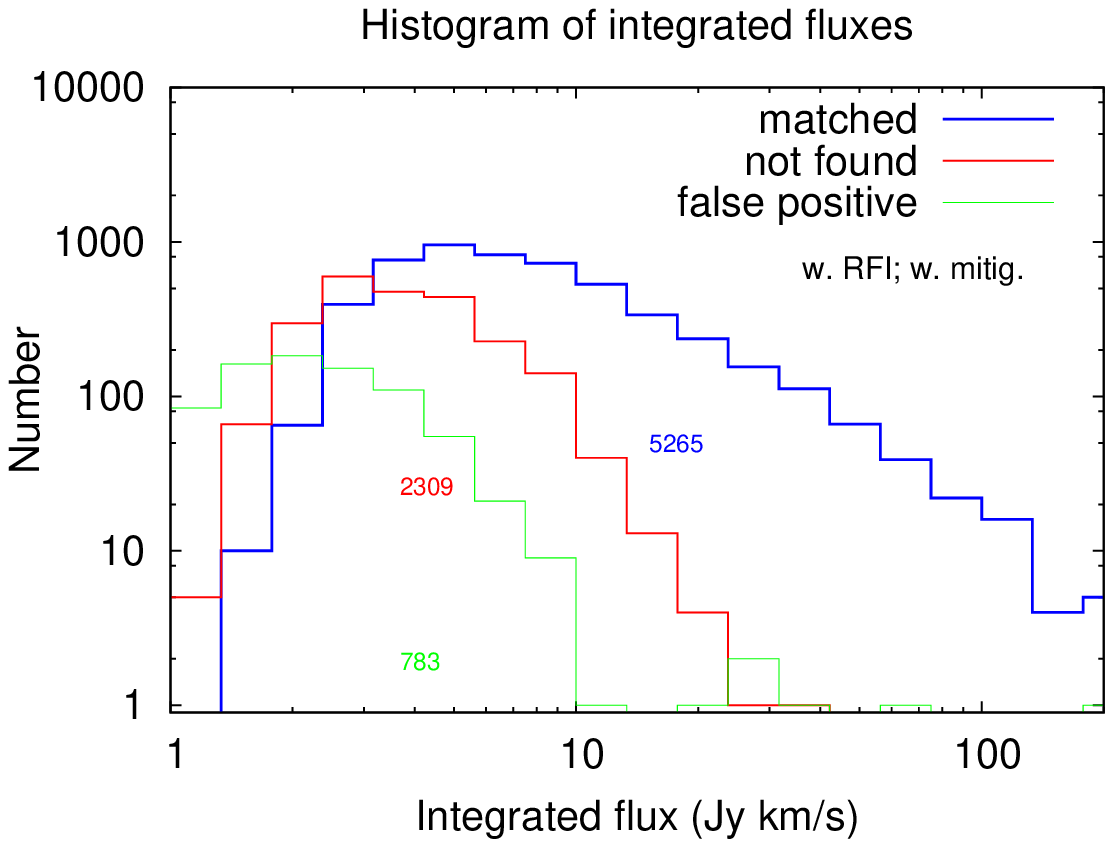}\raise8.5em\vbox{\moveleft11em\hbox{\small \it run (3)}}
\caption{Histograms of peak (top) and total (bottom) fluxes for all runs. Each plot shows the number of sources per peak flux interval for the correctly detected sources, \textit{false-positives}, and \textit{non-detections}. The colored numbers mark the total size of each list. }
\label{figanalysismatchedpeakandtotalfluxhisto}
\end{figure}

Fig.\,\ref{figanalysismatchedpeakandtotalfluxhisto} shows histograms of the derived peak (top panels) and total (bottom panels) fluxes for all runs. Each plot contains the numbers of sources per peak flux interval for the correctly detected sources, \textit{false-positives}, and \textit{non-detections}. As expected, the non-detections are low-flux sources. Except for \textit{run (2)} this also holds true for the false-positives. The latter resemble an overhead of about 15\% to 25\% compared to the correct detections (for \textit{run (2)} the number outreaches the matched galaxies by far). In \textit{run (2)} the number of matched galaxies decreased by a factor of three due to the high number of false-positives. The plots showing the integrated flux distributions are similar to those of the peak fluxes except for the false-positive detections, which inhibit a rather low total flux compared to the non-detections. The RFI signals in \textit{run (2)} are responsible for a second distribution of false-positives having very large total fluxes.
 One important result is, that the numbers of correct and non-detections in \textit{run (1)} and \textit{run (3)} are comparable, confining that the RFI mitigation works sufficiently well. 
Obviously, in the low-flux regime the number of non-detections is larger than the number of correct detections, while for higher fluxes the ``survey'' gets more and more complete. Generating sources below the actual detection limit, as well, is one way to determine the actual selection function.  Such a procedure was used to compute the completeness function for the HIPASS survey (Zwaan et~al. 2005 \cite{zwaan05}). 

\section{Systematic effects}\label{secsystematiceffects}
Some systematic effects in HIPASS were discussed by Zwaan et~al. (2004 \cite{zwaan04}). They compared generated and recovered parameters by computing the difference $\Delta P\equiv P^\mathrm{rec}-P^\mathrm{gen}$. To have a quantitative description, the resulting values $\Delta P$ were binned and Gaussians were fitted to the parameter distribution. The mean of the Gaussian, $\mu$, is an estimator for bias effects while the variance, $\sigma$, describes the scatter of the recovered quantities around the mean value. The recovered positional parameters (spatial coordinates and distance) do not show any deviation from   the generated values, however, the scatter increases towards smaller flux values. 

\begin{figure}[!t]
\centering
\includegraphics[scale=0.6,bb=20 57 240 458,clip=]{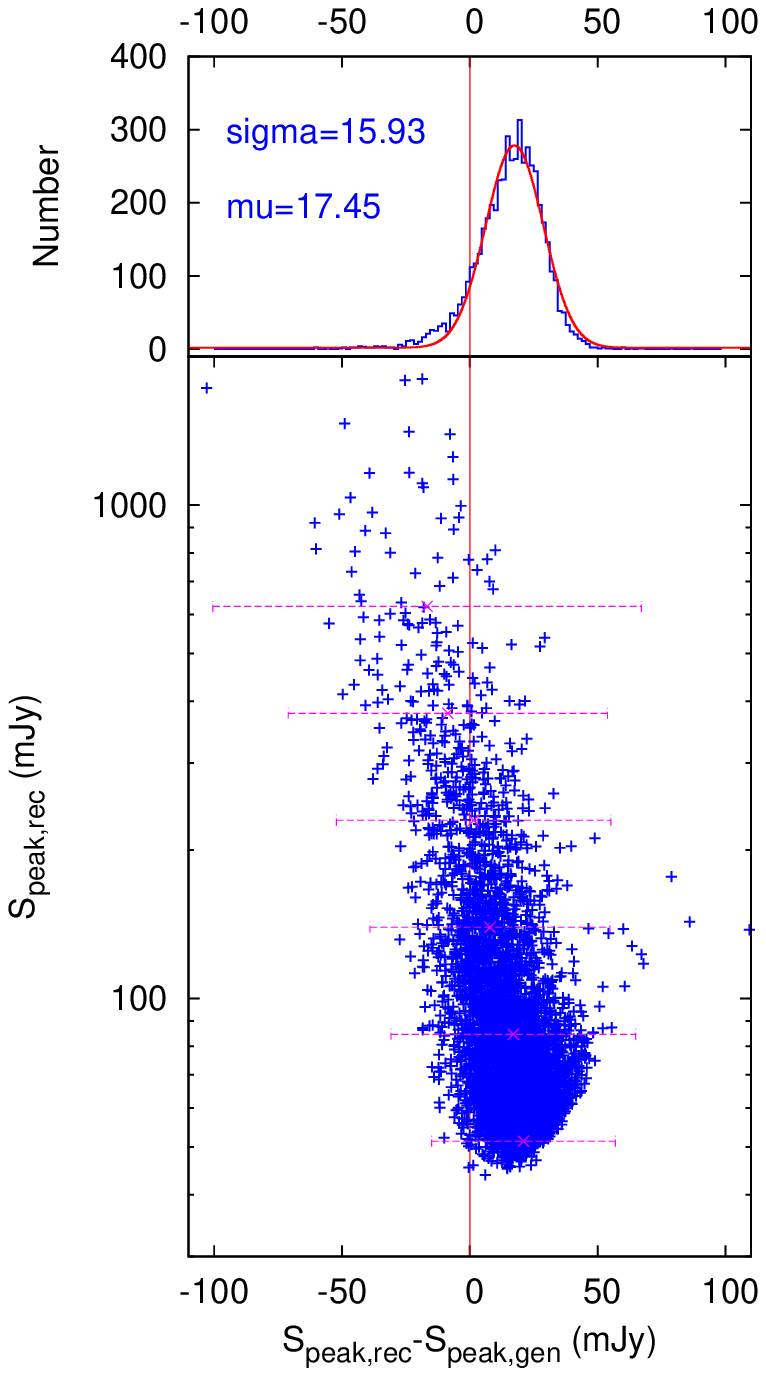}\quad
\includegraphics[scale=0.6,bb=20 57 240 458,clip=]{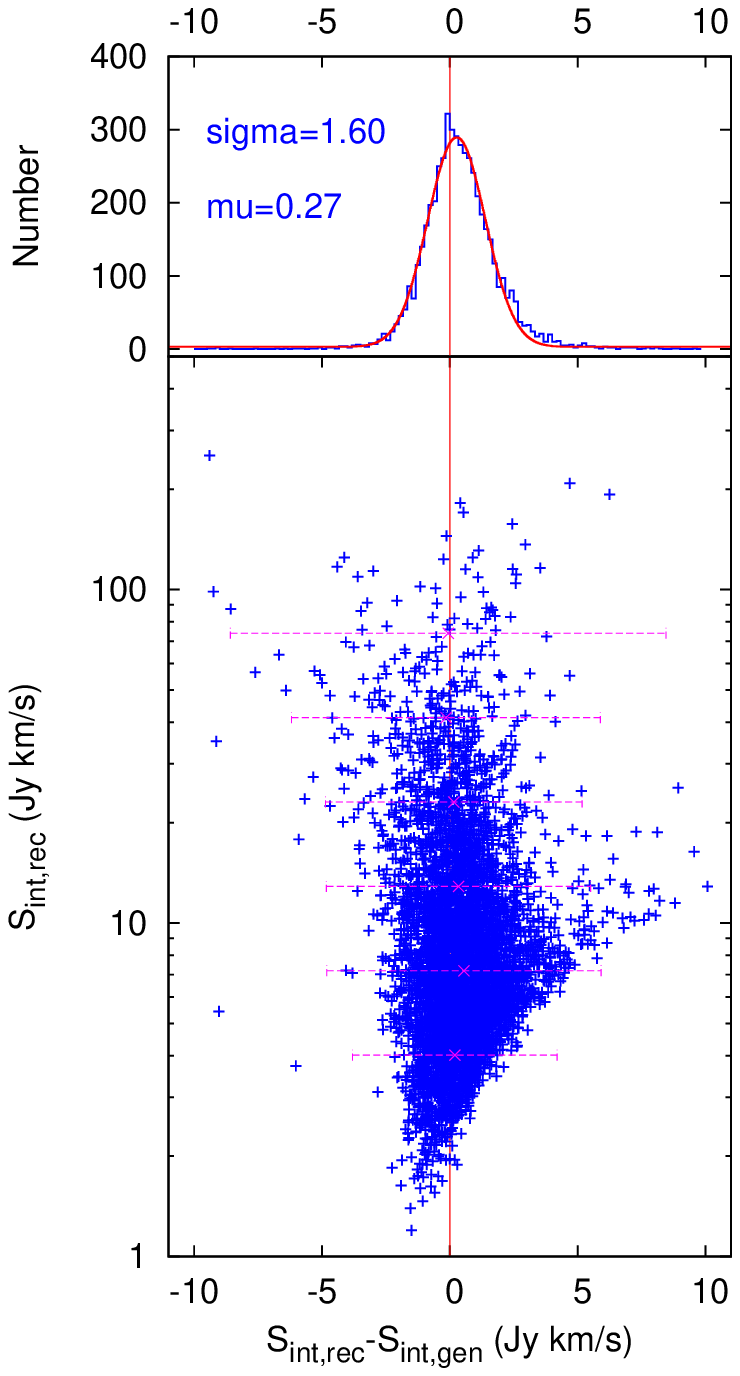}\quad
\includegraphics[scale=0.6,bb=20 57 240 458,clip=]{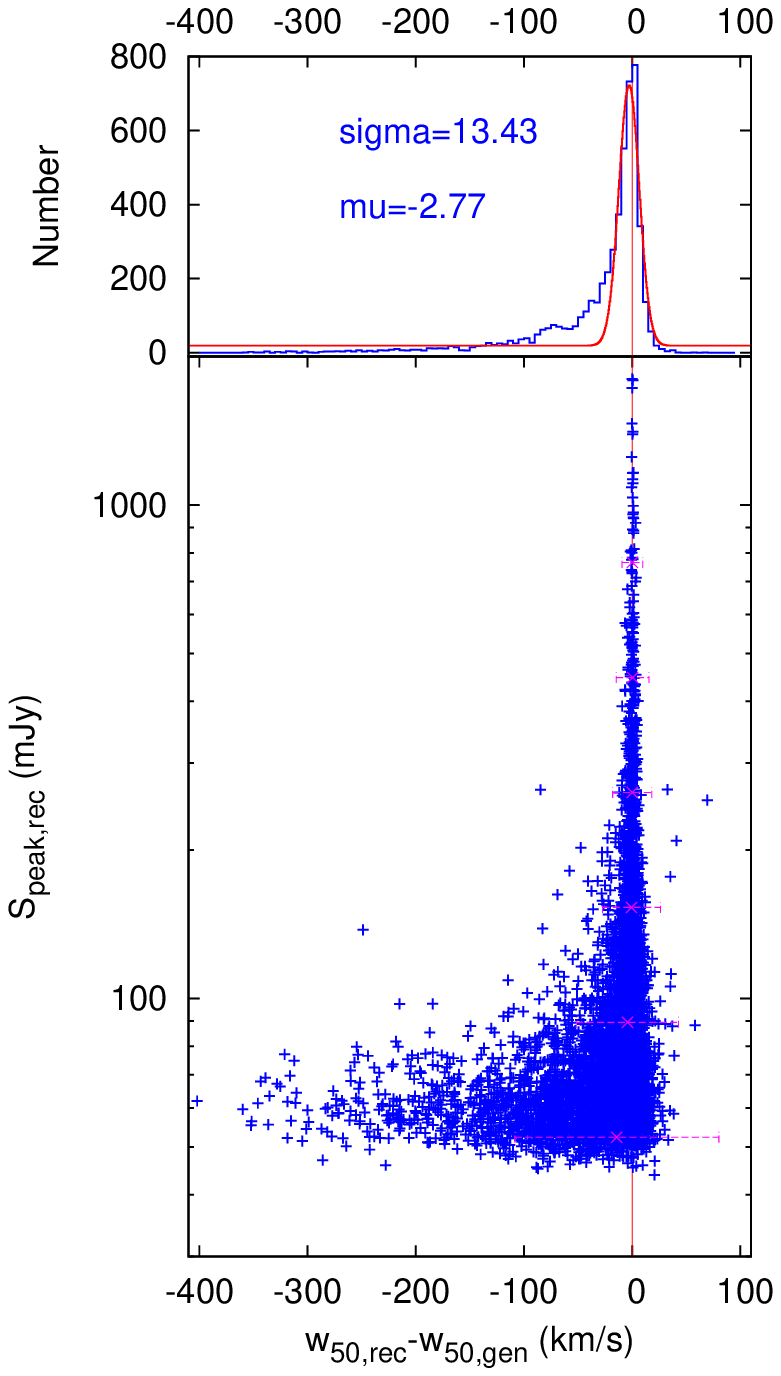}
\caption{Systematic differences in peak flux(left panel), integrated flux(middle panel), and velocity profile width (right panel) as a function of absolute peak and integrated flux, respectively.}
\label{figanalysissystematiceffectszwaanpeakandintflux}
\end{figure}

In Fig.\,\ref{figanalysissystematiceffectszwaanpeakandintflux} both, the differences between generated and recovered peak flux, $\Delta S_\mathrm{p}$ (left panel), and integrated flux, $\Delta S_\mathrm{int}$ (middle panel), respectively, are plotted as a function of reconstructed flux. The peak fluxes show a significant bias in the low-flux regime of $\mu\sim25\,\mathrm{mJy\,Beam^{-1}}$ as well as in the higher-flux regime ($\mu\sim-20\,\mathrm{mJy\,Beam^{-1}}$), counteracting each other. Furthermore, a slight overall shift of the recovered integrated fluxes to larger values is observed ($\mu=0.3\,\mathrm{Jy\,km/s}$). Additionally, there is a kind of ``wing'' for medium integrated flux values, produced by underestimation of $S_\mathrm{int}^\mathrm{rec}$ (for large values of $w_{50}$). In Fig.\,\ref{figanalysissystematiceffectszwaanpeakandintflux} (right panel) the velocity profile width, $w_{50}$,  deviations are plotted. They show a strong negative excess in the low flux regime. In the high-flux regime this bias is not seen. 

\begin{figure}[!t]
\centering
\includegraphics[scale=0.45,bb=75 73 396 294,clip=]{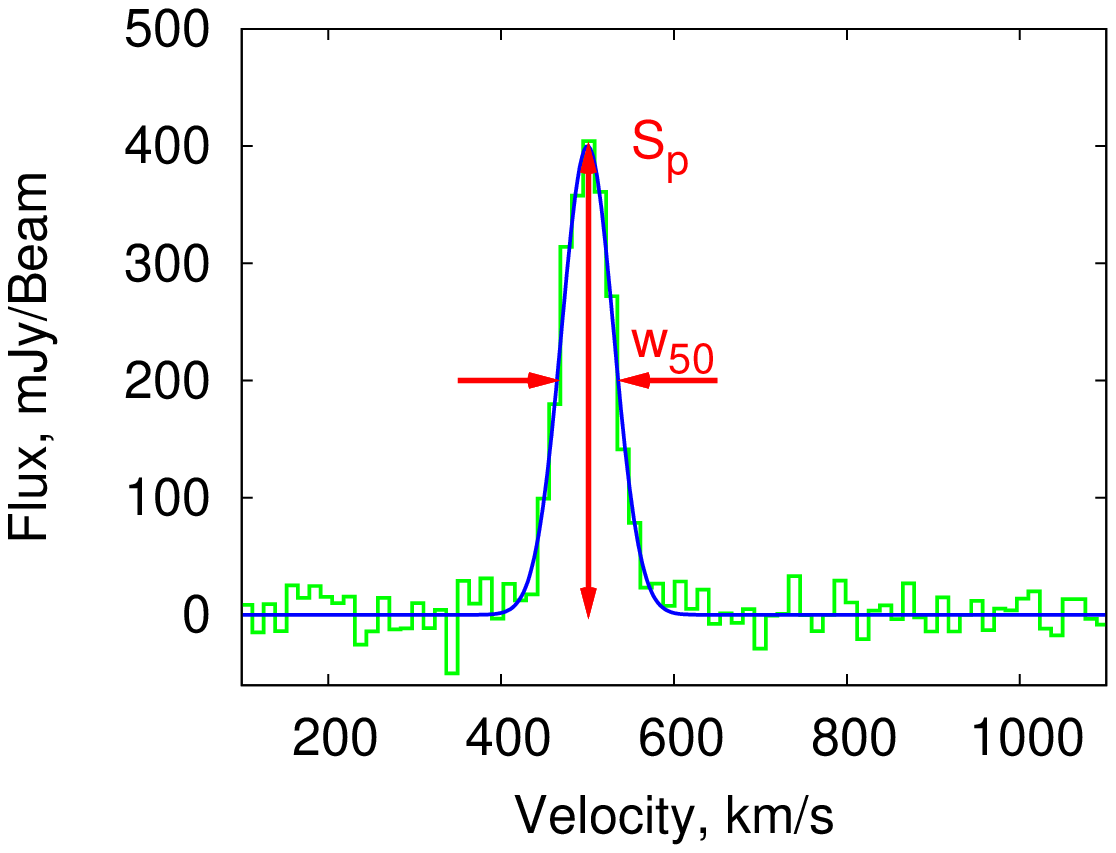}~
\includegraphics[scale=0.45,bb=95 73 396 294,clip=]{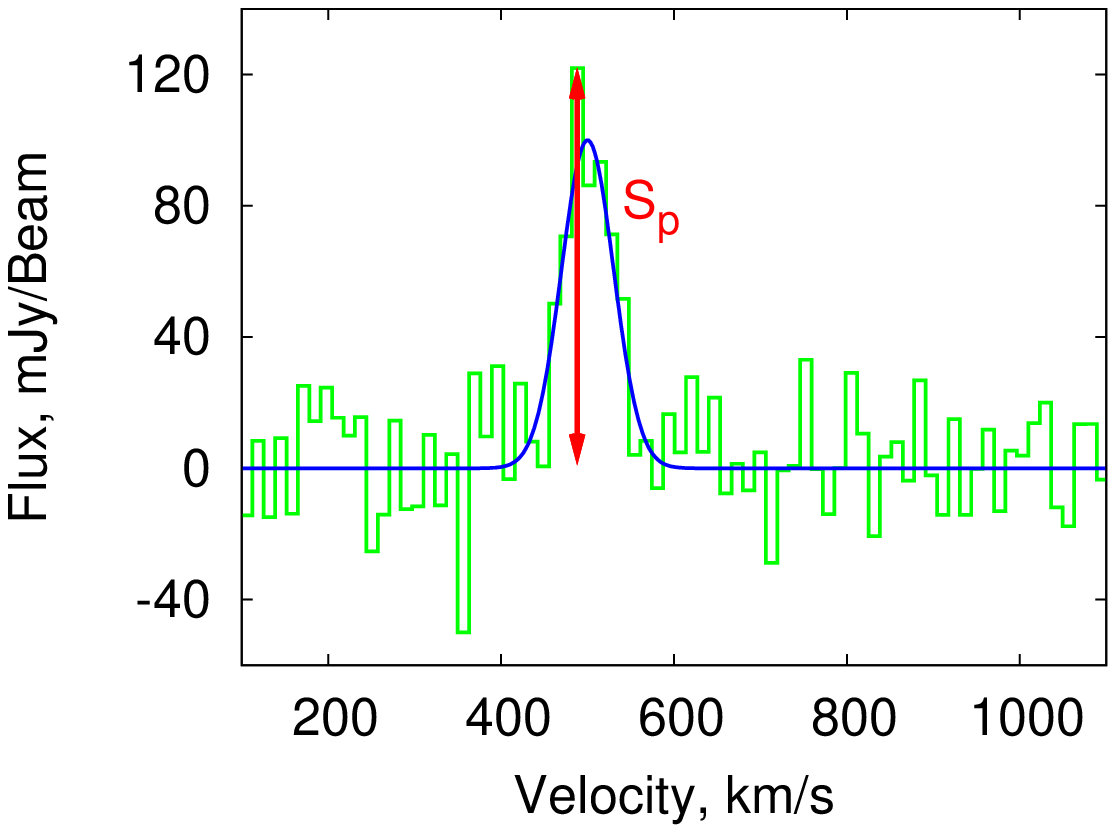}~
\includegraphics[scale=0.45,bb=95 73 396 294,clip=]{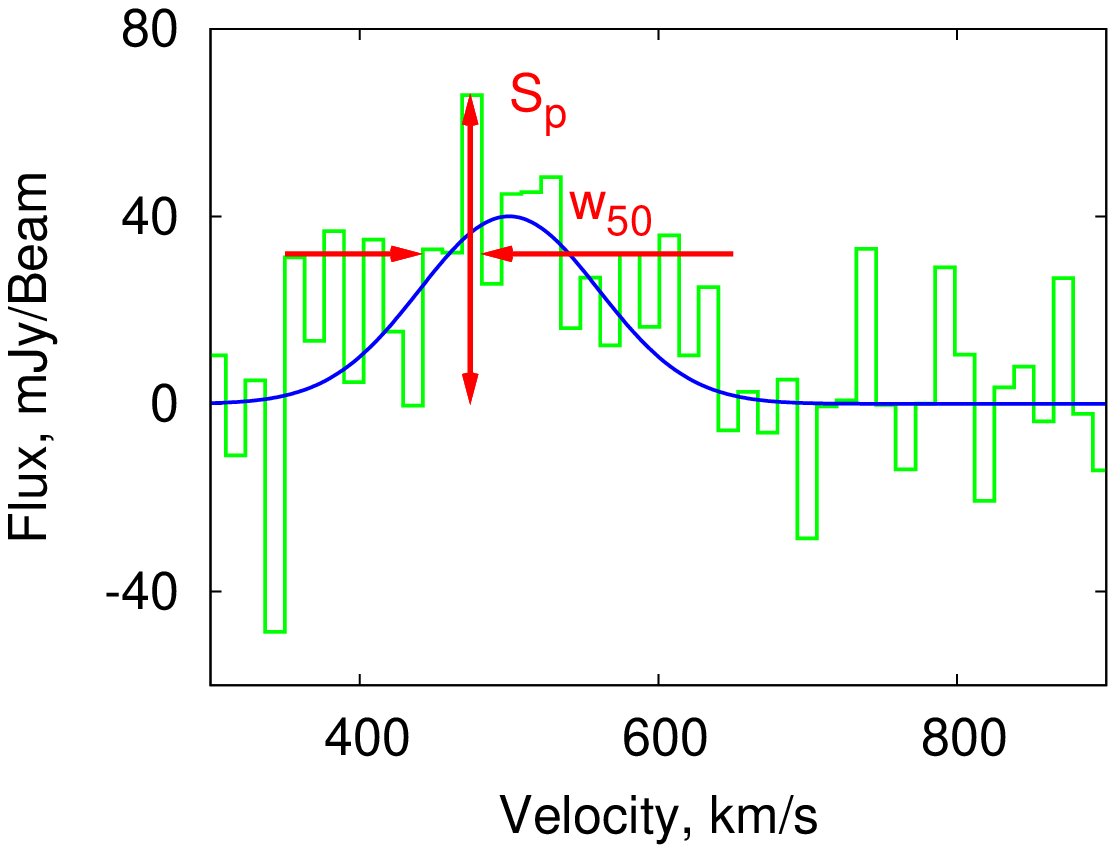}
\caption{\textbf{Left panel:} The parametrization used to obtain peak fluxes and velocity profile widths as used for the HIPASS catalog.  \textbf{Middle panel:} A single noise peak can add significantly to the underlying profile introducing a bias to the observed peak fluxes. \textbf{Right panel:} For low peak fluxes negative noise values can lead to a strong underestimation of the velocity width.}
\label{figanalysissystematiceffectsparametrization}
\end{figure}


The two effects seen in the peak fluxes are caused by gridding (which conserves the integrated flux but not necessarily the peak flux) and  the source parametrization method applied; see Fig.\,\ref{figanalysissystematiceffectsparametrization} (left panel). The peak flux is determined by finding the spectral channel with the highest flux in the velocity profile. This is not a good estimator in the presence of noise (see Fig.\,\ref{figanalysissystematiceffectsparametrization}, middle panel) and biases the peak fluxes to larger values, with a relatively stronger impact on smaller $S_\mathrm{p}^\mathrm{gen}$. Also, the overestimation of the width of the galaxy profile can be attributed to the parametrization (Fig.\,\ref{figanalysissystematiceffectsparametrization}, right panel). The width $w_{50}$ is calculated by searching for those spectral channels at which the spectral value has fallen below of 50\% the peak value. If the peak intensity is already low, it may happen, that these points lie below the noise and the width will be underestimated. Another source of uncertainty is due to a selection effect. From the completeness function follows, that objects with low total fluxes are found only to a certain fraction. Due to noise, a fraction of the sources will have a slightly increased detection probability (positive noise contribution). Hence, the faintest sources in the recovered sample are biased to higher flux values.


\end{document}